\documentclass[trackchanges, preprint, longbib]{aastex7}

\usepackage{amsmath} 
\submitjournal{PSJ}

\begin{document}


\title{Topographically Forced Stationary Waves Can Explain the Near-Surface Zonal Winds Observed by Huygens on Titan}

\author[orcid=0000-0003-1652-5365,sname='Han']{Sooman Han}
\affiliation{Department of Earth and Planetary Sciences, Yale University, New Haven, CT, USA}
\email[show]{sooman.han@yale.edu}
\correspondingauthor{Sooman Han}

\author[orcid=0000-0001-7464-1319,sname='Rafkin']{Scot Rafkin}
\affiliation{Department of Space Studies, Southwest Research Institute, Boulder, CO, USA}
\email{scot.rafkin@swri.org}

\author[orcid=0000-0003-3943-5194,sname='Battalio']{J. Michael Battalio}
\affiliation{Department of Earth and Planetary Sciences, Yale University, New Haven, CT, USA}
\email{joseph.battalio@yale.edu}

\author[orcid=0000-0002-7652-9758,sname='Fan']{Bowen Fan}
\affiliation{Department of Earth and Planetary Sciences, Yale University, New Haven, CT, USA}
\email{bowen.fan@yale.edu}

\author[orcid=0000-0001-9925-1050, sname='Lora']{Juan M. Lora} 
\affiliation{Department of Earth and Planetary Sciences, Yale University, New Haven, CT, USA}
\email{juan.lora@yale.edu}








\begin{abstract}
Titan’s equatorial superrotation is associated with climatological retrograde (westward) near-surface winds at low latitudes, as consistently simulated by Titan general circulation models. However, the Huygens probe measured prograde (eastward) near-surface winds, contrary to the expected circulation. We investigate whether these observations reflect localized forcing by large-scale topography, using paired three-dimensional dry simulations with and without topography in the Titan Atmospheric Model. We find that topography generates longitude-locked, vertically structured zonal wind anomalies that reproduce the observed features of the Huygens wind profile in the lowest few kilometers. Helmholtz and spectral analyses show that this structure arises from the superposition of several vertically varying low wavenumber modes. A predominantly rotational, topographically forced stationary zonal wavenumber $s=1$ response provides a major eastward anomaly throughout the lowest 2~km, whereas a locally divergent, $s=4$ response consistent with a gravity wave, opposes it near the surface. Weaker $s=2$ and $s=3$ responses also vary vertically and contribute eastward anomalies near the surface. Combined with the retrograde zonal-mean circulation, these modes produce a Huygens-like profile characterized by prograde flow within the lowest 1~km and shear toward retrograde flow aloft. The paired simulations establish their topographic origin, while coherence with the corresponding terrain-forcing components demonstrates that the responses remain phase locked to the forcing. Similar but weaker stationary responses are simulated near the planned Dragonfly landing site, providing baseline expectations for interpreting its future atmospheric measurements. This study demonstrates how large-scale topography can generate localized departures from Titan’s zonal-mean circulation and highlights the potential importance of stationary waves.
\end{abstract}




\section{Introduction} 
Titan, Saturn’s largest satellite, possesses a strongly superrotating equatorial atmosphere, with eastward stratospheric winds greatly exceeding the rotational speed of the solid body ($\sim$200 versus $10~\mathrm{m~s^{-1}}$) \citep{Achterberg08}. For solid planetary bodies, superrotation is favored in slowly rotating atmospheres and is maintained by the combined effects of the mean meridional circulation and eddy angular-momentum transport, as demonstrated in theoretical and modeling studies of Titan, Venus, and idealized atmospheres \citep[e.g.,][]{Gierasch75, DelGenio93, Lebonnois10, Mitchell&Vallis10, Newman11, Horinouchi20, Imamura20, Lombardo23b, Lewis23}. To maintain the requisite angular momentum balance, prograde winds at higher latitudes and at higher altitudes require the presence of climatological retrograde near-surface winds at low latitudes. Indeed, a range of Titan general circulation models (GCMs) consistently simulate retrograde (westward) mean near-surface winds in the equatorial region \citep{Newman11, Lebonnois12b, Lora15, Lora19}. 

In contrast, the near-surface winds observed by the Huygens probe appear inconsistent with this picture. Huygens landed at $10.3^\circ\mathrm{S}$, $167.7^\circ\mathrm{E}$ on January 14, 2005, during southern summer (solar longitude, $L_s\approx300^\circ$). The zonal wind profile measured using the Doppler Wind Experiment (DWE) indicated prograde (eastward) flow below approximately 1~km, overlain by retrograde flow between roughly 1 and 5~km \citep{Bird05, Folkner06}. An independent reconstruction of the probe trajectory further indicated southward meridional flow near the surface and northward flow aloft \citep{Karkoschka16}. Although this meridional wind reversal can be understood as part of Titan's seasonally varying cross-equatorial circulation \citep{Tokano07, Lora19, Lora24, Han&Lora25}, the observed near-surface prograde zonal flow remains difficult to reconcile with the retrograde equatorial near-surface winds simulated by Titan GCMs, suggesting that an additional mechanism is required.

Moreover, although surface friction can maintain vertical wind shear within a thermally convective planetary boundary layer (PBL), turbulent mixing tends to homogenize horizontal momentum and is not generally expected to produce a reversal in wind direction within the well-mixed layer \citep{Stull88, Garratt92, Lee16}. The observed wind reversal within the lowest few kilometers is therefore notable given that Titan’s PBL depth has been estimated to range from several hundred meters to approximately 2–3~km \citep{Tokano06b, Lorenz10a, Charnay12, Han&Lora25, Hartwick26}. Its coexistence with a near-adiabatic thermal structure suggests that processes beyond conventional PBL dynamics may have influenced the Huygens wind observations \citep{Tokano07}, although the responsible mechanisms remain unexplored.

Meanwhile, topographically forced stationary waves are generally recognized as important contributors to spatial variability in other planetary atmospheres \citep[e.g.,][]{Banfield03,Bertaux16}. A prominent example is the bow-shaped feature observed at the cloud tops of Venus, which has been attributed to vertically propagating gravity waves generated by underlying topography \citep{Fukuhara17,Lefevre20,Navarro24}. On Mars, interactions between wintertime zonal flow and large-scale topography excite stationary planetary Rossby waves, including prominent zonal wavenumber-1 and 2 disturbances that influence near-surface winds \citep{Hollingsworth96,Nayvelt97,Hinson03}.

Although we lack a directly measured, high-resolution global topographic map of Titan, Cassini RADAR provided elevation constraints through radar altimetry, SAR-derived topography, stereophotogrammetry, and local radarclinometry \citep[e.g.,][]{Lorenz13,Corlies17}. SAR imaging covered roughly $35\%$ of Titan’s surface, whereas the elevation measurements incorporated into the final mission-wide reconstruction covered about $9\%$. Elevations in unsampled regions were estimated through interpolation; for example, \citet{Corlies17} used radial basis functions to interpolate the expanded elevation dataset and construct a global topographic map. In this study, we investigate whether stationary waves forced by the global topographic reconstruction of \citet{Corlies17} help explain the zonal wind profile observed by Huygens within the lowest few kilometers. We compare otherwise identical Titan GCM simulations conducted with and without surface topography. Section~2 describes the model configurations and the observational data used in this study. Section~3 shows that, under the seasonal conditions corresponding to the Huygens descent, topography generates a persistent rotational response and a divergence-dominated, gravity wave-like response whose superposition reproduces key features of the observed wind profile. Section~4 discusses the implications for the Dragonfly mission, and Section~5 presents the conclusions.


\section{Methods}
\subsection{Titan Atmospheric Model}

We use the Titan Atmospheric Model (TAM), a three-dimensional Titan GCM that has been extensively evaluated against observations \citep{Lora15, Lora24}. To focus on the dynamical effects of topography, we employ a simplified dry configuration (i.e., without methane hydrologic cycle) with a spatially uniform, moderate surface thermal inertia of $I=548~\mathrm{J~m^{-2}~s^{-1/2}~K^{-1}}$, based on \citet{Han&Lora25}, and compare otherwise identical simulations with a flat surface and with topography. The topographic data are taken from the Cassini-based global reconstruction of \citet{Corlies17}, provided on a $0.25^\circ \times 0.25^\circ$ grid and interpolated onto the model’s T21 horizontal grid, which corresponds to $64\times32$ longitude--latitude points. The model employs 50 atmospheric levels, with a vertical resolution of several tens of meters within the lowest few kilometers, allowing boundary-layer-level variations to be resolved. Other configuration options are the same as in \citet{Han&Lora25}. Our analyses use a 40-Titan-day interval spanning $L_s=290^\circ$--$310^\circ$, drawn from the final simulated Titan year after a 250-Titan-year spin-up. Model output was sampled every $1/24$ Titan day ($\sim16$ Earth hours). This interval encompasses the seasonal conditions corresponding approximately to the Huygens descent and the anticipated arrival of Dragonfly.


\subsection{Huygens Observations}

We use the zonal wind profile retrieved from the Huygens DWE \citep{Bird05,Folkner06}, which extends from approximately 150~km altitude to the surface, although measurements are unavailable between roughly 5 and 12~km. The simulated zonal wind profiles in both experiments and the Huygens observations exhibit atmospheric superrotation. However, both simulations generally overestimate the observed eastward winds (Figure~A1), although superrotation is weaker in the topography simulation. The overestimation may partly reflect limitations of the dry-model configuration, which omits methane thermodynamics and moist convection. These processes can modify Titan’s meridional temperature gradients and thereby alter the vertical shear and strength of the zonal circulation through gradient-wind balance \citep{Mitchell09,Lora15,Newman16,ML16}. The weaker superrotation in the topography simulation may also result from topographic modification of wave propagation and angular momentum transport, both of which are important for Titan's superrotation \citep{Tokano12, Lewis23, Lian25}. In this study, we focus on the well-simulated wind structure within the lowest 5~km and do not investigate the upper-level differences further.


\section{Stationary Wave Analysis}

\subsection{Topographic Influence on the Zonal Wind Profile at the Huygens Landing Site}

Figure~1 compares zonal winds in the no-topography and topography simulations during southern midsummer ($L_s=290^\circ$--$310^\circ$) of the final simulated year. Figures~1(a,b) show Hovmöller diagrams of surface zonal wind at $8.3^\circ\mathrm{S}$, the model latitude nearest to the Huygens landing latitude of $10.3^\circ\mathrm{S}$. Relative to the no-topography simulation, the topography simulation exhibits a clear longitude-fixed structure, with persistently enhanced eastward winds near the Huygens longitude throughout the season. Figures~1(c,d) show seasonal-mean zonal winds in longitude--height sections. The topography simulation displays a vertically coherent wave structure relative to the underlying terrain, whereas the no-topography simulation is nearly uniform in longitude. Figures~1(e,f) compare the simulated wind profiles at the model grid point nearest the Huygens landing site $(163.1^\circ\mathrm{E},8.3^\circ\mathrm{S})$, with observations at the actual landing location $(167.7^\circ\mathrm{E},10.3^\circ\mathrm{S})$. In the topography simulation, prograde winds below approximately $1~\mathrm{km}$ are overlain by retrograde winds that reach their greatest magnitude near $3$--$4~\mathrm{km}$, followed by a return to prograde flow near $5~\mathrm{km}$. Most observed data points fall within the simulated minimum-to-maximum wind ranges. In contrast, zonal winds in the no-topography simulation vary more smoothly with height and remain retrograde throughout the lowest $5~\mathrm{km}$, failing to reproduce the observed winds within the first few hundred meters and near $4~\mathrm{km}$. These results show that topographically forced wind anomalies persist throughout the analyzed season and reproduce key features of the Huygens wind profile within the lowest few kilometers that are absent from the no-topography simulation.

\begin{figure}[htb!]
\centering
\includegraphics[width=0.84\textwidth]{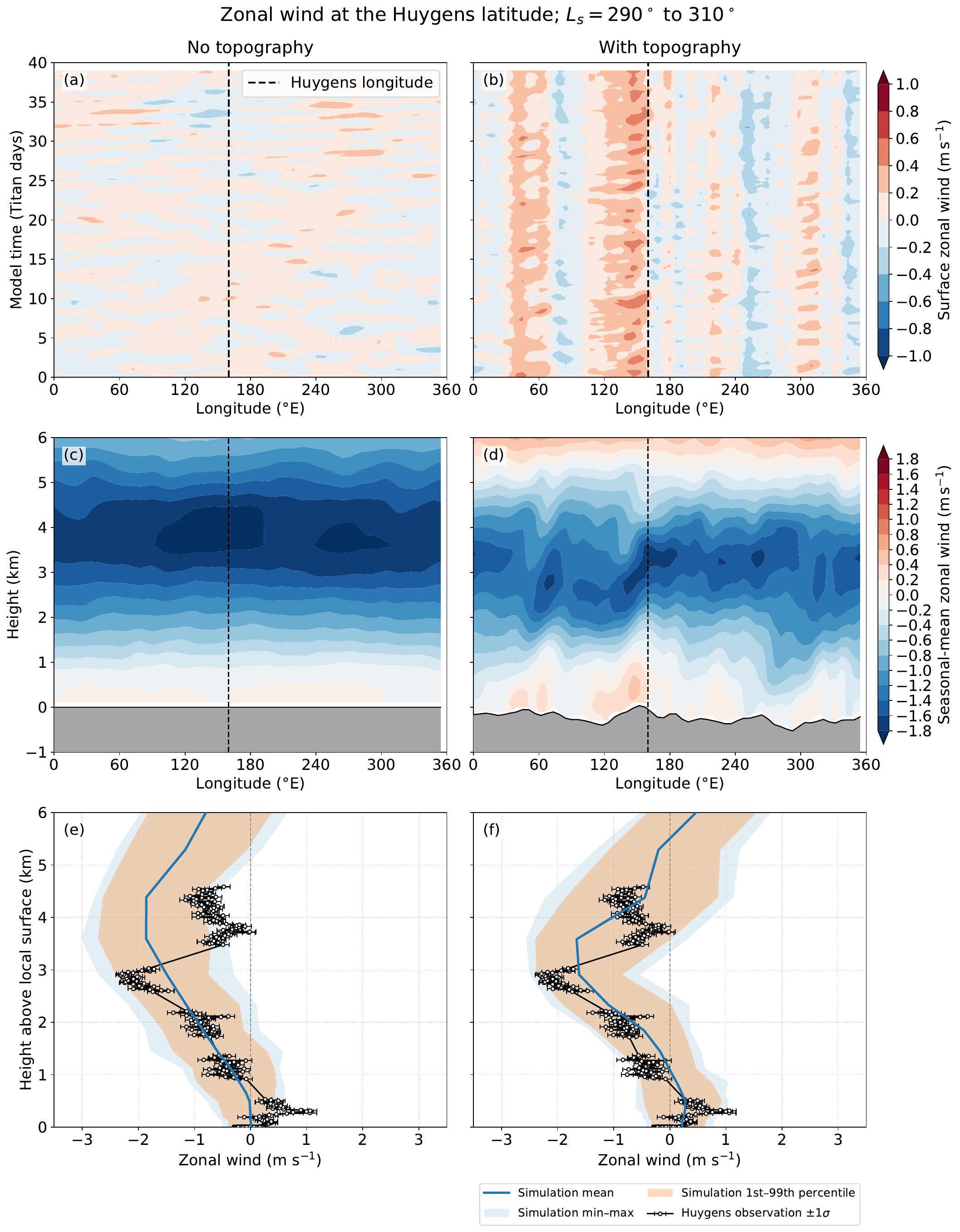}
\caption{Comparison of simulated zonal winds in the no-topography (i.e., flat-surface) and topography cases (left and right columns, respectively) during $L_s=290^\circ$--$310^\circ$ of the final year of the simulations. Panels (a) and (b) show Hovmöller diagrams of surface zonal wind at $8.3^\circ\mathrm{S}$, the model latitude nearest to the Huygens landing-site latitude of $10.3^\circ\mathrm{S}$. The landing longitude of $167.7^\circ\mathrm{E}$ (model grid point centered at $163.1^\circ\mathrm{E}$) is indicated by vertical dashed lines. Panels (c) and (d) show seasonal-mean zonal winds in longitude--height sections. Panels (e) and (f) compare the simulated zonal wind profiles at the landing site with the Huygens measurements.}
\end{figure}

To examine the topographic influence in greater detail, Figure~2 presents the seasonal-mean background zonal flow and stationary zonal-wind anomalies at the model latitude nearest the Huygens landing site. Figure~2(a) shows the zonal-mean wind profile, $\overline{U}(z)$ ($\mathrm{m\,s^{-1}}$), averaged over all longitudes and $L_s=290^\circ$--$310^\circ$ during the final simulated Titan year. The zonal-mean wind is approximately zero at the lowest model level but becomes retrograde immediately above and remains retrograde throughout most of the lowest $5~\mathrm{km}$. Although retrograde near-surface flow is expected in the long-term mean from angular-momentum constraints associated with atmospheric superrotation \citep{Gierasch75}, the near-zero surface wind during this interval reflects its seasonal variability. The prograde flow near the Huygens longitude in Figure~1(f) therefore represents a localized departure from the contemporaneous zonal-mean circulation rather than a reversal of the climatological near-surface flow. Figure~2(b) shows the corresponding vertical shear, $d\overline{U}/dz$ ($\mathrm{s^{-1}}$), which is negative below approximately $3~\mathrm{km}$, vanishes near the minimum in $\overline{U}$, and becomes positive above that altitude. Figure~2(c) shows a longitude--height section of the seasonal-mean stationary zonal wind anomaly, calculated by subtracting $\overline{U}(z)$ from the seasonal-mean wind at each longitude and height. Alternating positive and negative anomalies occur across longitude and exhibit coherent wave structure. At the Huygens longitude, marked by the dashed line, the anomaly is positive within approximately the lowest $2~\mathrm{km}$, becomes negative between roughly $2$ and $4~\mathrm{km}$, and turns positive again aloft. Figure~2(d) shows the spectrum of the topography at this latitude, demonstrating that its variance is concentrated primarily at low zonal wavenumbers, particularly $s=1$--$4$. Figure~2(e) shows the Fourier amplitude of the stationary zonal wind anomaly from Figure~2(c) as a function of zonal wavenumber and height, revealing particularly prominent and vertically extensive responses at $s=1$ and $s=4$. The corresponding height–wavenumber spectrum from the no-topography simulation shows no comparably prominent, vertically extensive $s=1$ or $s=4$ response (not shown). This contrast indicates that large-scale topographic forcing generates vertically structured stationary anomalies that produce localized departures from Titan’s zonal-mean circulation, including the prograde near-surface flow at the Huygens landing site.

\begin{figure}[htb!]
\centering
\includegraphics[width=0.66\textwidth]{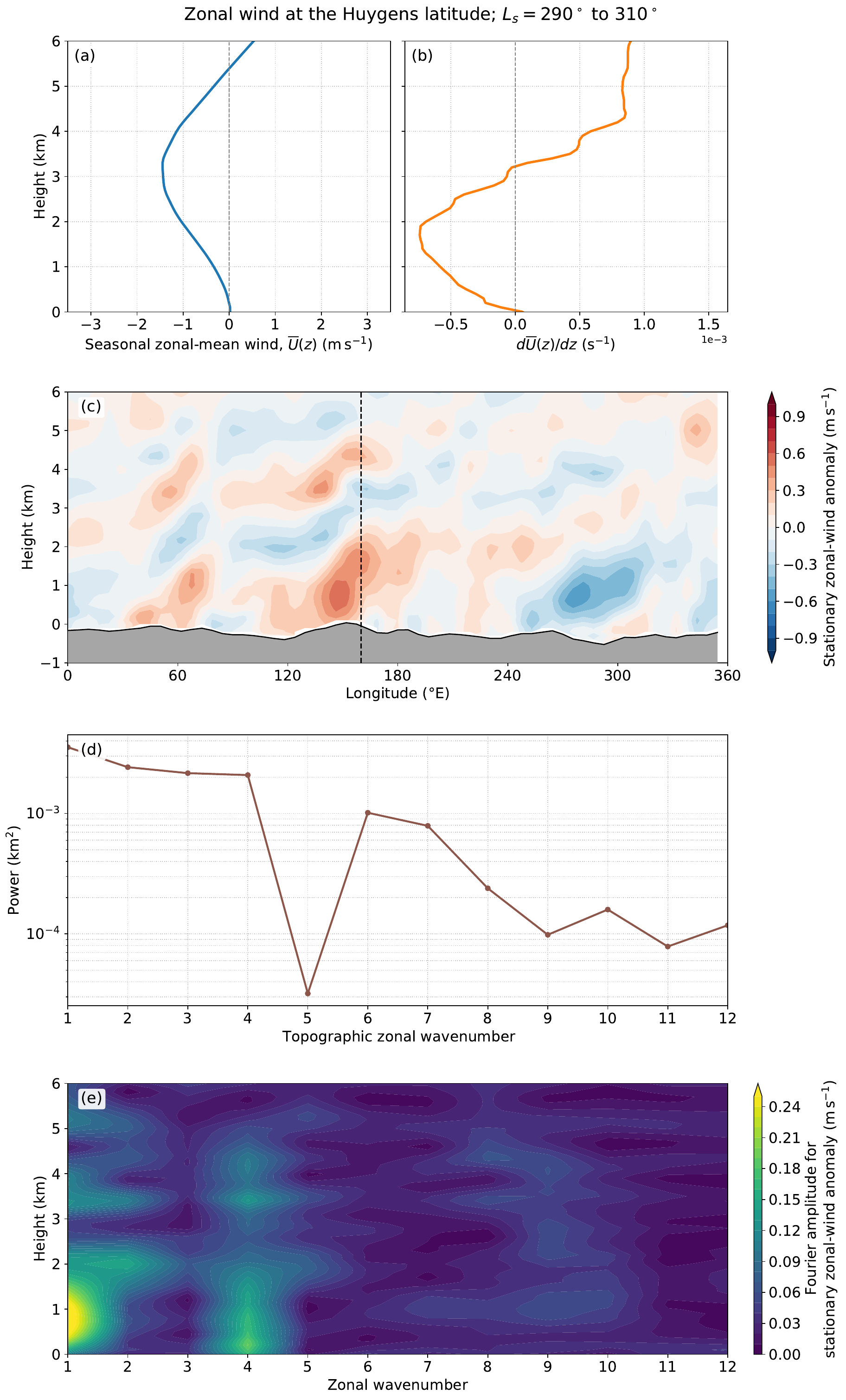}
\caption{Seasonal zonal-mean background flow and stationary wave structure at the model latitude nearest to the Huygens landing site ($\phi=-8.3^\circ$), in the topography simulation. (a) Seasonal zonal-mean zonal wind $\overline{U}(z)$ ($\mathrm{m\,s^{-1}}$). (b) Vertical shear $d\overline{U}/dz$ ($\mathrm{s^{-1}}$). (c) Longitude--height section of the seasonal-mean stationary zonal wind anomaly, obtained by subtracting $\overline{U}(z)$ from the seasonal-mean wind at each longitude and height. (d) Zonal wavenumber power spectrum of the topography. (e) Fourier amplitude of the stationary zonal wind anomaly in panel (c) as a function of zonal wavenumber and height.
}
\end{figure}

Longitude-fixed atmospheric disturbances can arise from geographically forced stationary Rossby waves or gravity waves \citep{Vallisbook17}. Motivated by the prominence of zonal wavenumbers $s=1$ and $s=4$ in Figure~2(e), we apply a horizontal Helmholtz decomposition on the sphere to the seasonal-mean stationary horizontal wind anomaly field \citep[e.g.,][]{Koshyk01, Hammond21} and evaluate the resulting rotational and divergent components at the Huygens landing site. Figure~3(a) shows that the $s=1$ anomaly is eastward within approximately the lowest $2~\mathrm{km}$ and closely follows the rotational component, whereas the divergent contribution is substantially smaller. The rotational component remains dominant aloft despite oscillations and sign reversals with height. In contrast, the $s=4$ anomaly is predominantly divergent throughout much of the lowest $4~\mathrm{km}$ and also exhibits pronounced vertical oscillations (Figure~3(d)). The weaker $s=2$ and $s=3$ modes provide additional eastward anomalies within the lowest few hundred meters (Figures~3(b,c)). Consequently, the combined contribution from $s=1$ through $s=4$ remains eastward near the surface because the eastward contributions from the $s=1$--$3$ modes collectively exceed the opposing $s=4$ contribution (Figure~3(e)). Thus, the simulated near-surface prograde flow arises from the combined influence of several low wavenumber stationary modes, while its broader vertical structure is dominated by the rotational $s=1$ response and the divergent $s=4$ response. In the following subsections, we assess whether these responses can be interpreted in terms of Rossby and gravity wave dynamics.

\begin{figure}[htb!]
\centering
\includegraphics[width=1.0\textwidth]{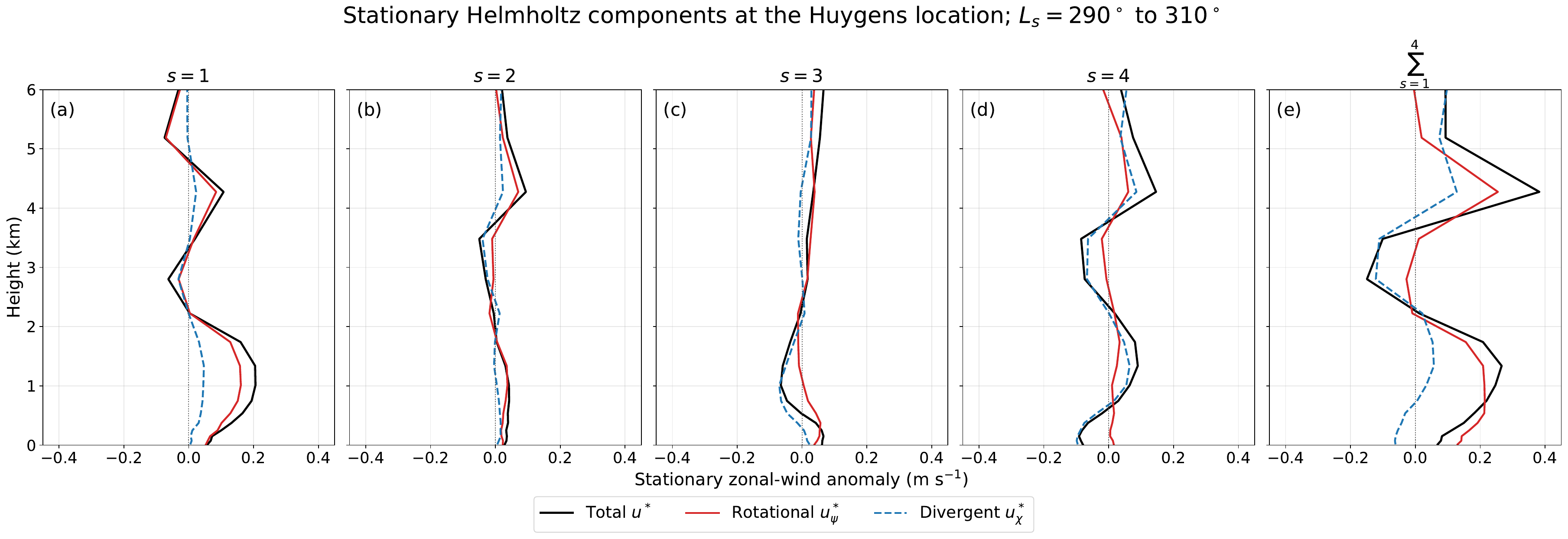}
\caption{
Vertical profiles of the Helmholtz components of the seasonal-mean stationary zonal wind anomaly in the topography simulation during $L_s=290^\circ$--$310^\circ$, evaluated at the model grid point nearest the Huygens landing site. Panels (a)–(d) show the contributions from zonal wavenumbers $s=1$ through $s=4$, respectively, and panel (e) shows their combined contribution. Red curves denote the rotational component $u_{\psi}^{*}$, blue dashed curves denote the divergent component $u_{\chi}^{*}$, and black curves denote their sum. Positive values indicate eastward flow, and the vertical dotted lines mark zero.
}
\end{figure}

\subsection{Interpretation of the Stationary $s=1$ Response}

A vertically propagating inertia--gravity wave interpretation of the $s=1$ component can be assessed using its intrinsic frequency. For a stationary disturbance, $\widehat{\omega}=-k\overline{U}$, where $\widehat{\omega}$ is the intrinsic angular frequency ($\mathrm{s^{-1}}$), $k=s/(a\cos\phi)$ is the physical zonal wavenumber ($\mathrm{m^{-1}}$), $a=2.58\times10^{6}$ is Titan’s mean radius (m), and $\phi$ is latitude. Vertical propagation requires $|f|<|\widehat{\omega}|<N$, where $f$ is the Coriolis parameter ($\mathrm{s^{-1}}$) and $N$ is the Brunt--Väisälä frequency ($\mathrm{s^{-1}}$). At the model grid point nearest Huygens ($\phi=-8.3^\circ$), $f=2\Omega\sin\phi$, with Titan’s rotation rate $\Omega=4.56\times10^{-6}~\mathrm{s^{-1}}$, gives $|f|=1.32\times10^{-6}~\mathrm{s^{-1}}$. For $s=1$, this corresponds to a lower propagation threshold of $U_{\mathrm{crit}}=|f|/k_{s=1}\approx3.4~\mathrm{m\,s^{-1}}$. Because the simulated background wind magnitude remains below this threshold throughout the analyzed layer (Figure~2(a)), the $s=1$ disturbance cannot be interpreted as a vertically propagating inertia--gravity wave within a linear framework.

Next, we assess a free Rossby wave interpretation. Under standard linear Rossby wave theory, the ground-frame frequency of a barotropic mode is

\begin{equation}
\omega=k\overline{U}-\frac{\beta k}{k^2+l^2},
\end{equation}
so a stationary mode $\omega=0$ requires

\begin{equation}
\overline{U}=\frac{\beta}{k^2+l^2}>0.
\end{equation}
Similarly, an equatorial Rossby mode satisfies

\begin{equation}
\omega=k\overline{U}
-\frac{\beta k}{k^2+(2n+1)/L_{\mathrm{eq}}^2},
\end{equation}
and stationarity requires

\begin{equation}
\overline{U}
=\frac{\beta}{k^2+(2n+1)/L_{\mathrm{eq}}^2}>0,
\end{equation}
where $k$ and $l$ are the physical zonal and meridional wavenumbers ($\mathrm{m^{-1}}$), $n$ is the equatorial meridional-mode index, $L_{\mathrm{eq}}$ is the equatorial deformation radius (m), and $\beta$ is the meridional gradient of the Coriolis parameter $(\mathrm{m^{-1}\,s^{-1}})$. Thus, a free Rossby mode can be stationary in the ground frame only under eastward background flow. The simulated retrograde flow therefore precludes interpreting the $s=1$ disturbance as a freely propagating stationary Rossby mode.

However, the distinct behavior of the no-topography and topography simulations (Figure~1(c,d)) indicates that these modes arise, at least in part, as forced responses rather than as free modes. Accordingly, we next consider whether the $s=1$ disturbance can be interpreted as a topographically forced stationary Rossby response. Geographically fixed $s=1$ topographic forcing imposes $\omega=0$. In the constant-density, quasi-geostrophic WKB approximation, the local squared vertical wavenumber, $m^2$ ($\mathrm{m^{-2}}$), of a stationary Rossby response is approximately \citep{Vallisbook17}

\begin{equation}
m^2=\frac{N^2}{f^2}
\left[
\frac{\beta}{\overline{U}}
-\left(k^2+l^2\right)
\right].
\end{equation}
Because $\beta>0$, westward background flow ($\overline{U}<0$) gives $m^2<0$, implying a vertically evanescent or trapped response in this simplified quasi-geostrophic framework. Such a forced response may exhibit standing baroclinic structure, including vertical nodes and discrete phase reversals, but is not expected to exhibit systematic vertical phase progression through regions of appreciable amplitude. In contrast, the longitude–height structure of the simulated $s=1$ rotational zonal wind anomaly exhibits systematic vertical phase tilt, rather than only discrete phase reversals across amplitude nodes (Figure~A2), which does not clearly support a vertically trapped Rossby response. We therefore interpret the $s=1$ disturbance conservatively as a topographically forced rotational response of an inconclusive nature, rather than specifically as a vertically propagating inertia--gravity wave, a free stationary Rossby mode, or a vertically trapped Rossby response.

\subsection{Interpretation of the Stationary $s=4$ Response}


The locally dominant divergent component of the $s=4$ response at the Huygens landing site (Figure~3(d)) motivates a linear gravity wave analysis. Figure~4 therefore examines whether the vertical structure and polarization of the $s=4$ disturbance are consistent with linear gravity wave theory \citep{Durran90,Vallisbook17}. As noted above, vertical propagation of an inertia--gravity wave requires $|\overline{U}|>U_{\mathrm{crit}}=|f|/k_{s=4}$. At the model latitude nearest Huygens, this gives $U_{\mathrm{crit}}\approx0.84~\mathrm{m\,s^{-1}}$ for $s=4$. Because the magnitude of the simulated background wind is comparable to this threshold, particularly below 2~km and above 4~km, the $s=4$ disturbance lies near the inertial propagation cutoff in these layers. Rotation therefore cannot be neglected, and the following analysis based on nonrotating gravity wave polarization relations should be regarded as a qualitative check on consistency rather than an exact quantitative prediction.

Figures~4(a--c) show the longitude--height structures of the $s=4$ zonal wind, vertical velocity, and potential temperature anomalies, respectively. The zonal wind and vertical velocity anomalies exhibit coherent eastward phase tilts (Figures~4(a,b)). Above approximately $1~\mathrm{km}$, the potential temperature anomaly exhibits a broadly similar phase tilt (Figure~4(c)), whereas its stronger near-surface component displays a distinct structure that is discussed in the following subsection. The zonal wind anomaly weakens near 5~km, where the background zonal wind approaches zero, corresponding to a critical level for a stationary wave because $\widehat{\omega}=-k\overline{U}\rightarrow0$ (Figure~2(a)). The vertical velocity anomaly begins to weaken at approximately 3 to 4~km, near the altitude where the vertical shear changes sign and the propagation condition $|\overline{U}|>U_{\mathrm{crit}}$ is no longer satisfied (Figures~2(a,b)). Although this is not a critical level because $\overline{U}$ remains nonzero, it may represent an inertial-cutoff region in which vertical propagation is inhibited.

The overall phase structure is qualitatively consistent with upward gravity wave propagation. In the nonrotating hydrostatic limit, the vertical wavenumber $m$ ($\mathrm{m^{-1}}$) satisfies $m\approx N/\overline{U}$. Because a line of constant phase satisfies $kx+mz=\mathrm{constant}$, its phase slope is

\begin{equation}
\frac{dx}{dz}
=
-\frac{m}{k}
\approx
-\frac{N}{k\overline{U}},
\end{equation}

\noindent where $x$ is eastward distance and $z$ is height. Because the background flow is westward ($\overline{U}<0$), this relation gives $dx/dz>0$. The simulated eastward phase tilt with height is therefore qualitatively consistent with an upward-propagating stationary gravity wave.

\begin{figure}[htb!]
\centering
\includegraphics[width=0.93\textwidth]{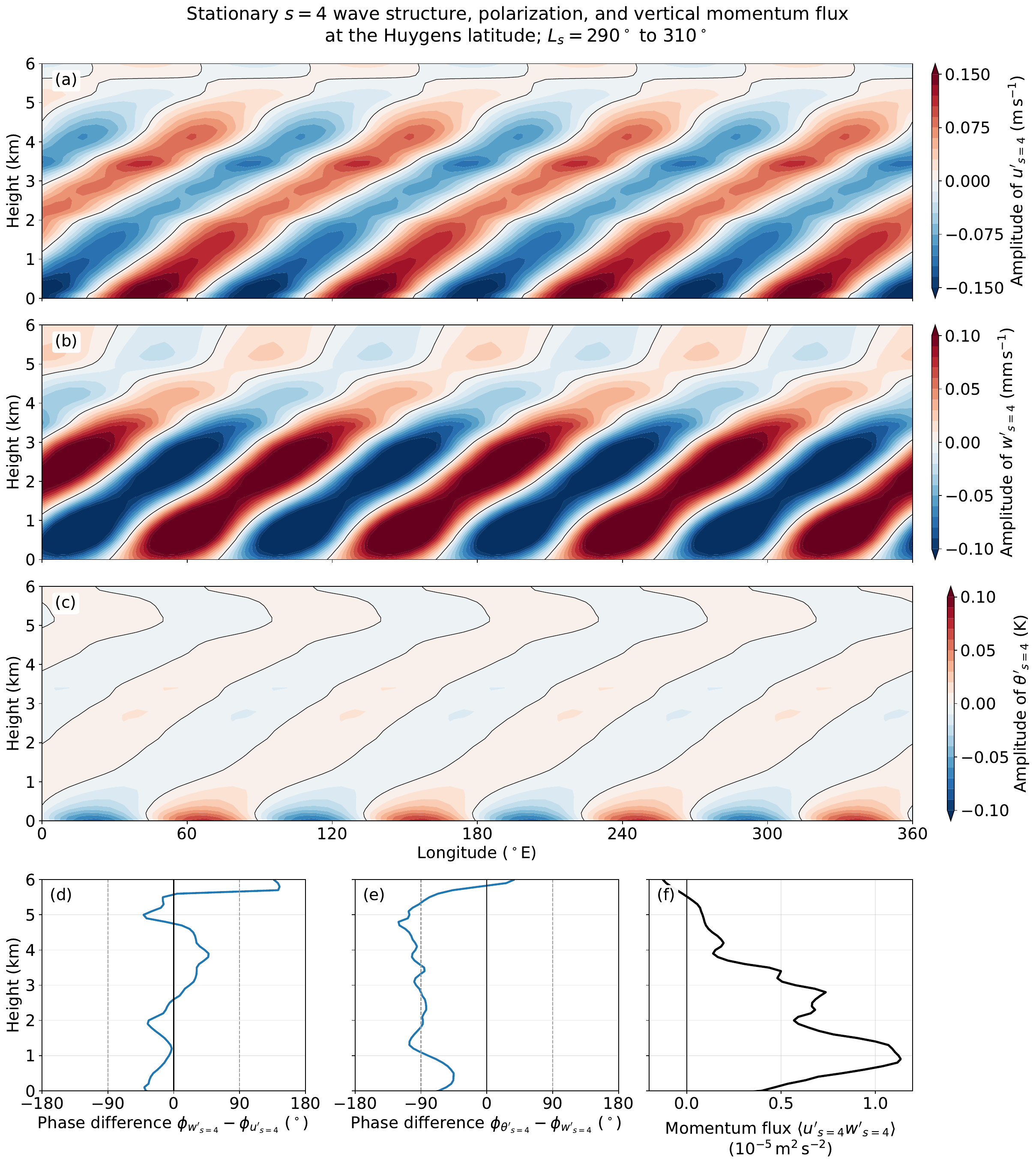}
\caption{
Vertical structure, polarization, and momentum transport of the stationary $s=4$ disturbance in the topography simulation during $L_s=290^\circ$--$310^\circ$, at the model latitude nearest to the Huygens landing site. Panels (a)--(c) show longitude--height sections of the reconstructed $s=4$ zonal wind anomaly $u'_{s=4}$ ($\mathrm{m\,s^{-1}}$), vertical velocity anomaly $w'_{s=4}$ ($\mathrm{mm\,s^{-1}}$), and potential temperature anomaly
$\theta'_{s=4}$ (K), respectively. Panels (d) and (e) show the vertical profiles of the phase differences
$\phi_{w'_{s=4}}-\phi_{u'_{s=4}}$ and $\phi_{\theta'_{s=4}}-\phi_{w'_{s=4}}$, respectively. Solid vertical lines indicate $0^\circ$, and dashed lines indicate $\pm90^\circ$. Panel (f) shows the vertical momentum flux $\langle u'_{s=4}w'_{s=4}\rangle$, where angle brackets denote a longitudinal average; positive values indicate upward transport of eastward momentum.
}
\end{figure}

The polarization diagnostics provide further support for the gravity wave interpretation. The phase difference between $w'_{s=4}$ and $u'_{s=4}$ remains relatively close to $0^\circ$ through most of the lowest $5~\mathrm{km}$ (Figure~4(d)), indicating that the two velocity components are predominantly in phase. Meanwhile, $\theta'_{s=4}$ generally lags $w'_{s=4}$ by approximately $90^\circ$ between about $1$ and $5~\mathrm{km}$ (Figure~4(e)), consistent with the thermodynamic polarization expected for a linear gravity wave in a stably stratified atmosphere. Below approximately $1~\mathrm{km}$, the phase relationships depart more substantially from these linear predictions. This deviation likely arises because the background zonal wind is weak near the surface (Figure~2(a)), such that the small-amplitude condition $|u'|/|\overline{U}|\ll1$ required by linear theory is not satisfied. Moreover, this region lies within Titan’s diurnally evolving PBL \citep{Han&Lora25}, where dry convection and turbulent mixing can modify the phase relationships predicted for waves propagating through a steady, stably stratified background. The deviations below $1~\mathrm{km}$ therefore likely reflect a breakdown of the linear gravity wave assumptions.

The vertical momentum flux, $\langle u'_{s=4}w'_{s=4}\rangle$, is predominantly positive from the surface to approximately 5~km, peaks near 1~km, and generally decreases with height above that level (Figure~4(f)). The increase up to 1~km is not expected under steady, conservative linear wave dynamics. Instead, as noted above, it likely reflects nonconservative processes within Titan’s diurnally evolving PBL where dry convection and turbulent mixing can modify the wave and prevent conservation of its momentum flux. Because the background zonal flow is retrograde, this positive momentum flux is consistent with upward energy propagation under the approximate stationary wave relation $F_z=\langle p'_{s=4}w'_{s=4}\rangle \approx -\rho_0U\langle u'_{s=4}w'_{s=4}\rangle$. However, we reiterate that rotation cannot be neglected for this disturbance so the relation again simply provides a qualitative consistency test. Nevertheless, together, the coherent phase tilt, gravity wave-like polarization, and nonzero vertical momentum flux support a topographically forced gravity wave-like response whose phase structure is consistent with upward propagation within the 2 to 4~km layer where $|\overline{U}|>U_{\mathrm{crit}}$. The disturbance is inertially evanescent below approximately 2~km, near its lower-boundary source; it may tunnel into this limited propagation window and become evanescent again above approximately 4~km as the background wind weakens toward the critical-level region. Determining whether its subsequent weakening results from dissipation, absorption, or interference with reflected waves is beyond the scope of this study.

\subsection{Horizontal Structure and Terrain-Forcing Relationships}

To examine the broader spatial organization of the responses discussed above, Figures~5 and~6 show the horizontal structures of the $s=1$ and $s=4$ components from Figure~3 at $z\approx0.54~\mathrm{km}$, where the stationary wind anomalies are pronounced, across all longitudes and from $30^\circ\mathrm{S}$ to $30^\circ\mathrm{N}$. Figure~5(a) shows that the $s=1$  relative vorticity and stationary wind anomaly fields form broad, planetary-scale circulation cells. Relative vorticity at the Huygens site is negative and therefore cyclonic in the Southern Hemisphere, corresponding to clockwise rotation. The streamfunction and rotational wind reveal a coherent gyre in the southern tropics (Figure~5(b)), with Huygens situated on its equatorward flank, where the rotational zonal flow is eastward. Although an eastward divergent contribution is also present at the Huygens longitude (Figure~5(c)), the longitudinal profile at the Huygens latitude shows that the total eastward anomaly at the landing site is dominated by the rotational component (Figure~5(d)). The near-surface $s=1$ anomaly is therefore embedded within a large-scale vortical circulation, further supporting its interpretation as a large-scale, topographically forced rotational response.

\begin{figure}[htb!]
\centering
\includegraphics[width=1.0\textwidth]{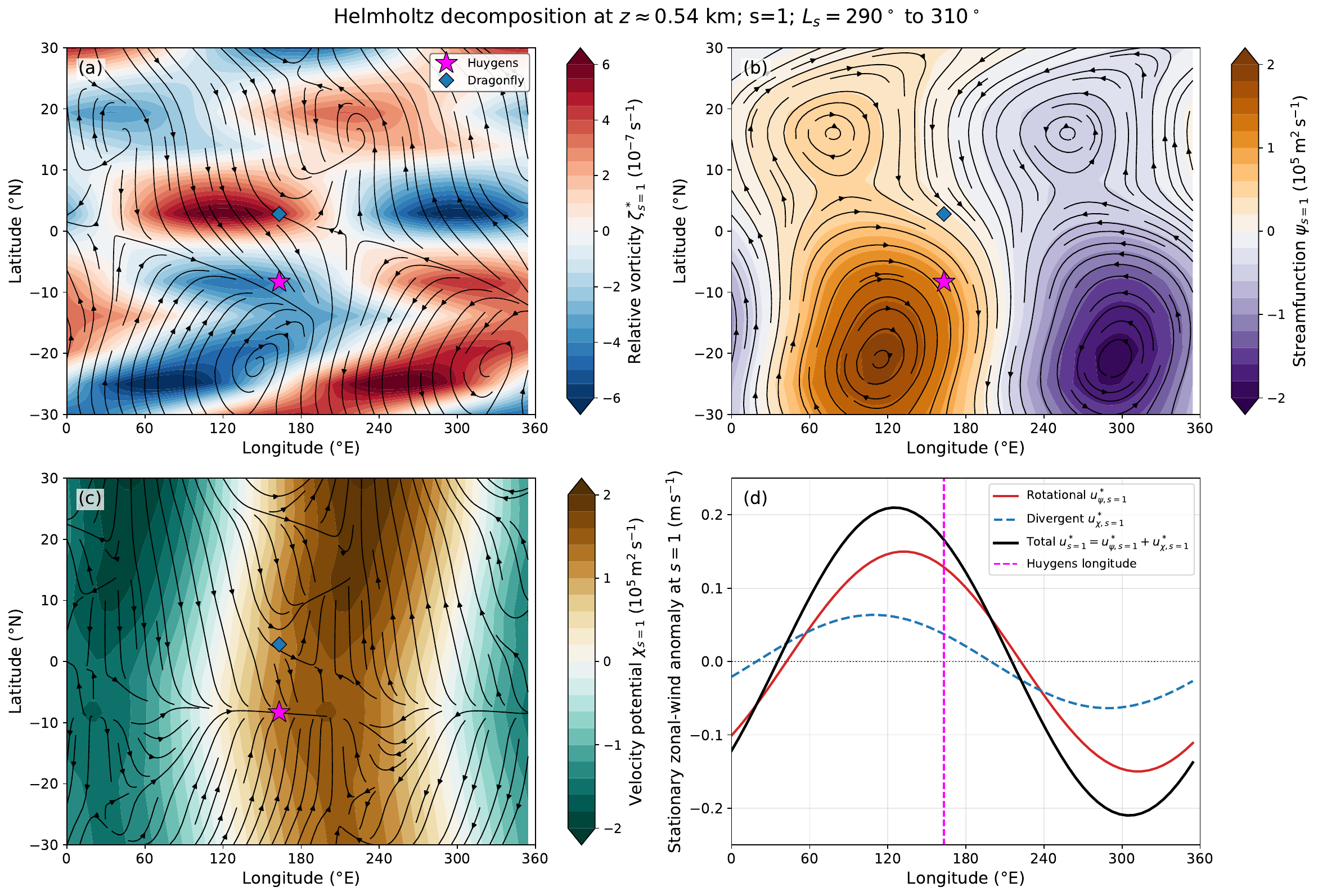}
\caption{
Horizontal structure of the seasonal-mean stationary $s=1$ wind anomaly and its Helmholtz components at $z\approx0.54~\mathrm{km}$ during $L_s=290^\circ$--$310^\circ$. (a) Relative vorticity (shading) and total $s=1$ wind anomaly streamlines. (b) Streamfunction (shading) and rotational wind streamlines. (c) Velocity potential (shading) and divergent wind streamlines. The magenta star marks the model grid point nearest to the Huygens landing site, and the blue diamond marks the Dragonfly latitude at the same model longitude. (d) Longitudinal profiles of the total (black), rotational (red), and divergent (blue dashed) zonal wind anomalies at the Huygens latitude. The magenta vertical dashed line marks the Huygens longitude, and positive values denote eastward flow.
}
\end{figure}

Figure~6(a) shows that the $s=4$  relative vorticity and stationary wind anomaly fields form a shorter wavelength pattern comprising four repeating circulation structures around the globe. As for the $s=1$ component, the Huygens landing site lies within a region of negative relative vorticity. The streamfunction and rotational wind components reveal a sequence of smaller-scale gyres (Figure~6(b)), while the velocity potential and divergent wind exhibit a pronounced alternating pattern (Figure~6(c)). Figure~6(d) shows that the rotational zonal wind component generally has the larger amplitude across most longitudes. At the Huygens longitude, however, the rotational contribution is relatively weak, whereas the divergent component dominates locally and produces a westward total anomaly. The near-surface $s=4$ response therefore partially offsets the eastward $s=1$ contribution. Thus, although the global $s=4$ pattern contains a substantial rotational circulation, its expression at Huygens is predominantly divergent.

\begin{figure}[htb!]
\centering
\includegraphics[width=1.0\textwidth]{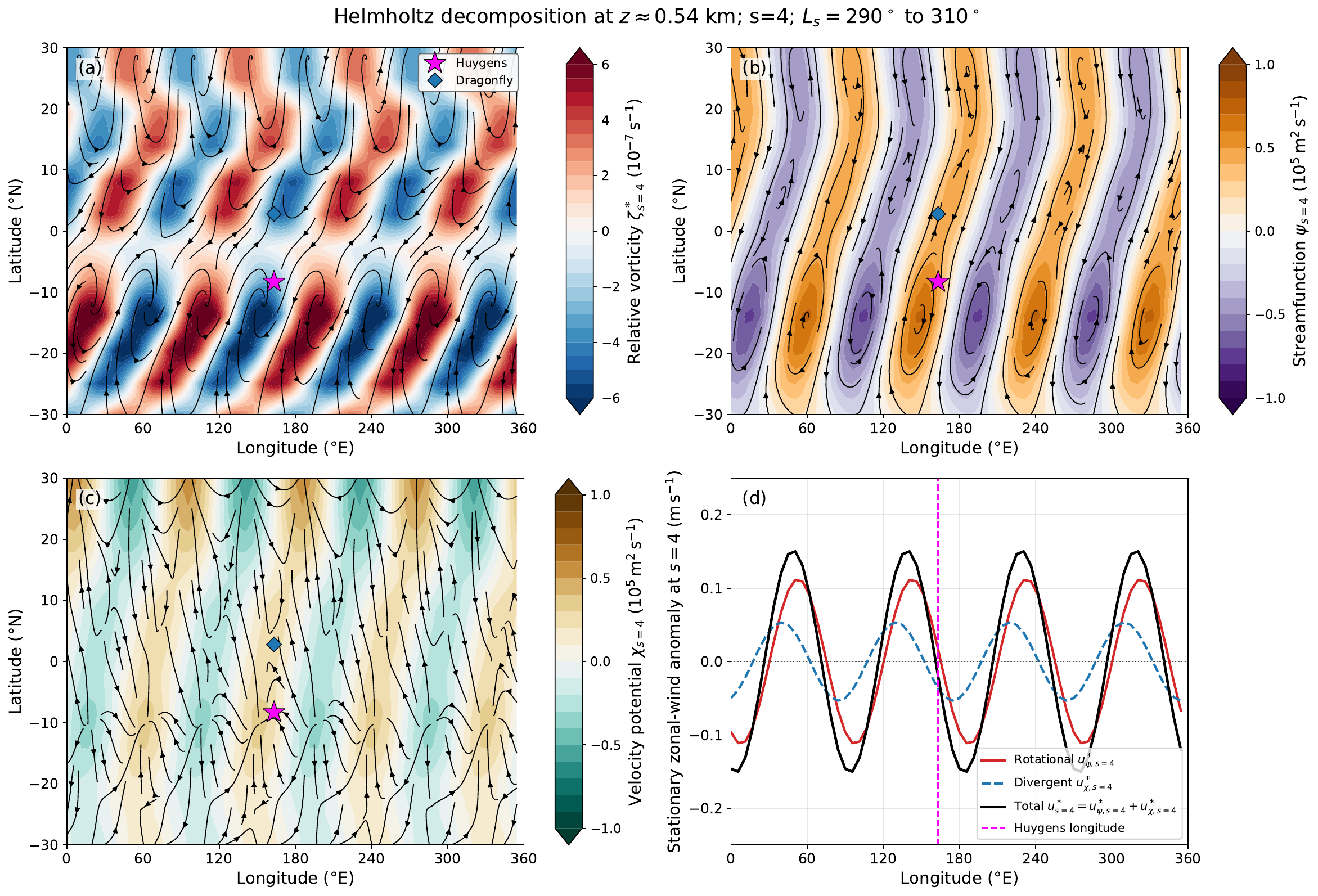}
\caption{
Same as Figure~5, but for the stationary $s=4$ component.
}
\end{figure}

The comparison between the topography and no-topography simulations establishes that the prominent stationary modes arise from topographic forcing (Figure~1). To characterize how these modes are dynamically coupled to the lower boundary, Figure~7 compares the $s=1$ and $s=4$ atmospheric responses with the corresponding kinematic forcing generated by the background surface flow crossing topographic slopes. This forcing is defined using the linearized kinematic lower-boundary
condition \citep[e.g.,][]{Smith79,Durran90} as

\begin{equation}
w_h
=
\overline{\mathbf{U}}_{\mathrm{sfc}}\boldsymbol{\cdot}\nabla h
=
\frac{\overline{U}_{\mathrm{sfc}}}{a\cos\phi}
\frac{\partial h}{\partial\lambda}
+
\frac{\overline{V}_{\mathrm{sfc}}}{a}
\frac{\partial h}{\partial\phi}.
\end{equation}

\noindent where $h$ is the surface elevation (m), $\lambda$ is east longitude, and $\overline{\mathbf{U}}_{\mathrm{sfc}}=(\overline{U}_{\mathrm{sfc}},\overline{V}_{\mathrm{sfc}})$ is the zonal-mean surface wind ($\mathrm{m\,s^{-1}}$). The terrain-forcing velocity, $w_h$, represents the vertical velocity ($\mathrm{m\,s^{-1}}$) required for the background flow to follow the terrain: positive values indicate upslope ascent and negative values indicate downslope descent. We first calculate $w_h$ and isolate each zonal wavenumber $s$. At each latitude, we then compare the complex Fourier coefficient of the terrain forcing, $W_{s,j}$, with that of each atmospheric response, $R_{s,j}$, where $R$ denotes either relative vorticity or horizontal divergence, and $j$ indexes eight nonoverlapping five-Titan-day blocks spanning $L_s=290^\circ$--$310^\circ$. Sensitivity tests using different block lengths and numbers of blocks produce qualitatively similar coherence values (not shown). This comparison uses the magnitude-squared coherence \citep[e.g.,][]{Bendat10}, defined as

\begin{equation}
\gamma_s^2 =
\frac{
\left|
\left\langle
R_{s,j} W_{s,j}^{*}
\right\rangle_j
\right|^2
}{
\left\langle |R_{s,j}|^2 \right\rangle_j
\left\langle |W_{s,j}|^2 \right\rangle_j
},
\end{equation}

\noindent which measures the consistency of the combined amplitude-and-phase relationship between the forcing and response. It ranges from zero to one, with values near one indicating a reproducible amplitude-and-phase relationship between the forcing and response across the eight blocks.

Figures~7(a,b) show broad planetary-scale $s=1$ topography and four alternating $s=4$ ridge-and-trough structures. Because $w_h=\overline{\mathbf{U}}_{\mathrm{sfc}}\boldsymbol{\cdot}\nabla h$ depends on topographic slopes rather than elevation, the forcing and atmospheric responses need not align with elevation extrema. At Huygens, westward flow produces positive $s=1$ forcing on the eastern flank of an elevation anomaly, whereas the $s=4$ zonal forcing is weak near the elevation extremum. On the other hand, comparison of Figures~4(c) and~7(b) indicates that the near-surface $s=4$ potential temperature anomaly is approximately in phase with the topography, with positive $\theta'$ over elevated terrain and negative $\theta'$ over topographic troughs. This relationship is opposite to that expected from adiabatic displacement in a stably stratified atmosphere, which would produce negative $\theta'$ over topographic crests. Instead, it supports a surface thermal-boundary mechanism: for an approximately uniform surface temperature, the lower pressure over elevated terrain produces higher potential temperature. The expected anomaly, $(g/c_p)h\approx0.13~\mathrm{K}$ for $h\approx100~\mathrm{m}$, is comparable to the simulated near-surface amplitude of approximately $0.1~\mathrm{K}$ in Figure~4(c).

The magnitude-squared coherence between each atmospheric response and the corresponding terrain-forcing component, $w_{h,s}$, calculated using Eq.~(8) and shown in Figures~7(c,d), provides a consistency check that the modes are phase-locked to the terrain-forcing component. Under the null hypothesis of zero true coherence, the magnitude-squared coherence estimated from $K$ independent spectral realizations follows a $\mathrm{\beta}(1,K-1)$ distribution. The corresponding 95\% significance threshold is $\gamma^2_{95}=1-0.05^{1/(K-1)}$, which gives $\gamma^2_{95}\approx0.35$ for $K=8$. Thus, magnitude-squared coherence exceeding 0.35 would occur by chance in fewer than 5\% of realizations if the forcing and response were unrelated. Note that this threshold is nominal because the blocks contain the same stationary component and may not be statistically independent; their effective number of independent estimates may therefore be smaller than eight, making the actual threshold higher. At the Huygens latitude, the magnitude-squared coherence values for relative vorticity and divergence are 0.89 and 0.95, respectively, for $s=1$, and 0.91 and 0.96 for $s=4$. These values, all substantially above the nominal threshold, indicate reproducible amplitude-and-phase relationships across the eight blocks.

The cross-spectral phases provide additional information about the spatial relationship between the forcing and response (not shown). For $s=1$, the vorticity and divergence phases relative to $w_{h,s=1}$ are $127.6^\circ$ and $99.4^\circ$, corresponding to response offsets of approximately $127.6^\circ$ and $99.4^\circ$ west of the forcing, respectively. For $s=4$, the corresponding phases are $52.6^\circ$ and $64.1^\circ$, equivalent to westward offsets of approximately $13.2^\circ$ and $16.0^\circ$. Thus, both modes are phase-locked to the terrain-forcing component but spatially displaced from it, with a substantially larger response–forcing offset for $s=1$.

\begin{figure}[htb!]
\centering
\includegraphics[width=1.0\textwidth]{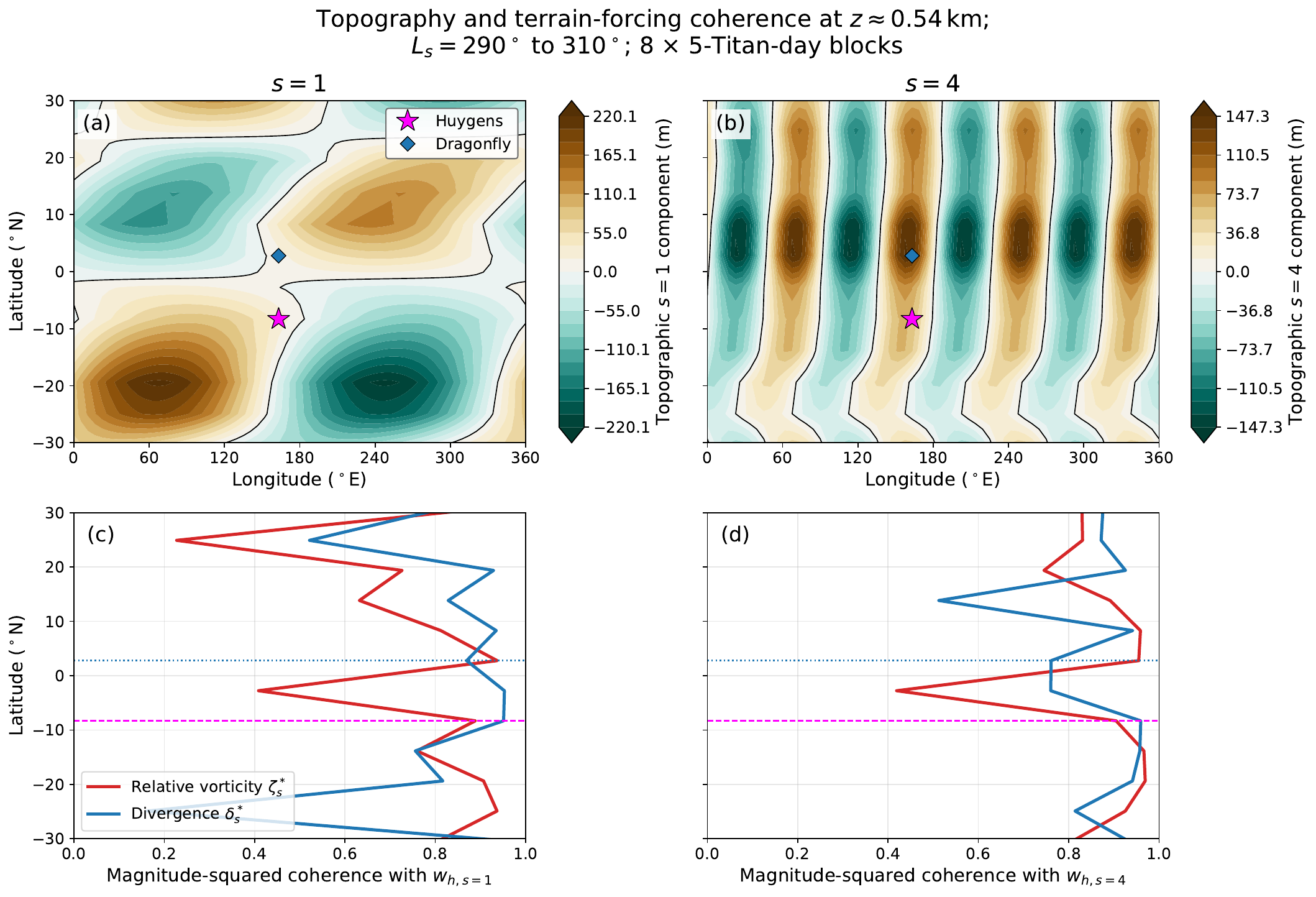}
\caption{
Topographic structure and terrain-forcing coherence for the stationary $s=1$ and $s=4$ modes at $z\approx0.54~\mathrm{km}$ during $L_s=290^\circ$--$310^\circ$. Panels (a) and (b) show the $s=1$ and
$s=4$ Fourier components of surface elevation, respectively. The magenta star marks the model grid point nearest the Huygens landing site, and the blue diamond marks the Dragonfly latitude at the same model longitude.
Panels (c) and (d) show the magnitude-squared coherence of relative vorticity (red) and horizontal divergence (blue) with the corresponding kinematic terrain forcing,
$w_{h,s}=(\overline{\mathbf{U}}_{\mathrm{sfc}}\boldsymbol{\cdot}
\nabla h)_s$, calculated across eight nonoverlapping five-Titan-day blocks using Eq.~(8). The magenta dashed and blue dotted lines indicate the Huygens model latitude and Dragonfly latitude, respectively.
}
\end{figure}

\section{Expectations for Dragonfly}

In this section, we use the same topographic simulation data to derive testable predictions for NASA’s future Dragonfly mission \citep{Barnes21,Lorenz21a}. Dragonfly is expected to arrive during a season similar to that of the Huygens descent ($L_s\approx309^\circ$) and at a nearby longitude, but in the opposite hemisphere. Its planned landing location is approximately $(161.8^\circ\mathrm{E},3.7^\circ\mathrm{N})$, represented in the model by the nearest grid point at $(163.1^\circ\mathrm{E},2.8^\circ\mathrm{N})$. Thus, the principal geographic difference between the two sites is their position on opposite sides of the equator: Dragonfly will be in the northern winter hemisphere, whereas Huygens descended in the southern summer hemisphere. Figures~5(b,c) and 6(b,c) show that the two locations lie within the same positive streamfunction and velocity-potential lobes for both the $s=1$ and $s=4$ modes. This correspondence suggests that the stationary disturbances retain a similar longitudinal phase across the equator and that Dragonfly may encounter rotational and divergent structures related to those identified near Huygens.

Figure~A3 repeats the analysis of Figure~2 at the model latitude nearest the planned Dragonfly landing site. As at the Huygens latitude, the seasonal zonal-mean wind is retrograde throughout approximately the lowest 5~km before becoming prograde aloft (Figure~A3(a)). The corresponding vertical shear is slightly more negative near the surface than at the Huygens latitude, remains predominantly negative below approximately 3~km, and becomes positive above the minimum in the background wind (Figure~A3(b)). At the Dragonfly longitude, the stationary zonal wind anomaly is predominantly positive within approximately the lowest $1.5~\mathrm{km}$, negative between about $1.5$ and $3~\mathrm{km}$, and positive again between roughly $3$ and $4.5~\mathrm{km}$ (Figure~A3(c)). The topographic power spectrum remains dominated by low zonal wavenumbers, with particularly strong contributions from $s=2$ and $s=4$ and a smaller $s=1$ component (Figure~A3(d)). The atmospheric response nevertheless exhibits prominent $s=1$ and $s=4$ structures (Figure~A3(e)), although the $s=1$ amplitude is weaker than at the Huygens latitude, while the near-surface $s=4$ response is comparatively pronounced. Thus, the same two stationary modes remain relevant at Dragonfly, but their relative contributions differ from those at Huygens, with a greater relative role for the shorter wavelength $s=4$ disturbance.

Figure~A4 applies the Helmholtz decomposition at the model grid point nearest the planned Dragonfly landing site. The $s=1$ anomaly is predominantly eastward throughout approximately the lowest $1.5~\mathrm{km}$, with its near-surface maximum controlled mainly by the rotational component and a smaller contribution from the divergent component (Figure~A4(a)). In contrast, the $s=4$ response is predominantly divergent, with a secondary rotational contribution (Figure~A4(d)); its westward near-surface anomaly partially offsets the eastward $s=1$ response before alternating between eastward and westward anomalies aloft. Figure~A4(b,c) shows that the $s=2$ and $s=3$ modes produce westward and eastward anomalies, respectively, within the lowest few hundred meters, with both reversing sign with height. Dragonfly may therefore encounter the same qualitative superposition identified at Huygens—a rotational $s=1$ circulation combined with a more divergent, gravity wave-like $s=4$ disturbance and weaker, vertically varying $s=2$ and $s=3$ contributions—but with different relative amplitudes and vertical phases (Figure~A4(e)).

Figure~A5 provides further evidence that the $s=4$ response at the Dragonfly latitude is gravity wave-like. The $s=4$ anomalies in zonal wind, vertical velocity, and potential temperature exhibit coherent vertical phase tilts between approximately 1 and $4~\mathrm{km}$ altitude. Over much of this layer, $u'_{s=4}$ and $w'_{s=4}$ are nearly in phase, while $\theta'_{s=4}$ lags $w'_{s=4}$ by approximately $90^\circ$, broadly consistent with linear gravity wave polarization. The predominantly positive vertical momentum flux below approximately $4~\mathrm{km}$ further indicates upward energy propagation under the retrograde background flow. Thus, Dragonfly may encounter a vertically propagating, topographically forced $s=4$ gravity wave disturbance similar to that identified near Huygens.

Finally, Figure~8 repeats the analysis of Figure~1 at the model latitude nearest the planned Dragonfly landing site. In the no-topography simulation, the seasonal-mean near-surface flow is retrograde and contains an approximately $200$-m-thick layer of nearly uniform wind, coincident with the shallow near-surface adiabatic layer simulated during winter by \citet{Han&Lora25}. With topography, this layer shifts toward more positive zonal winds, although its seasonal-mean velocity remains close to zero. Topography also produces more eastward flow near the surface and above approximately $3$--$4~\mathrm{km}$, consistent with the stationary wave structures identified in Figure~A4. These topographically induced departures are weaker at Dragonfly than at Huygens, potentially reflecting differences in both hemispheric background circulation and local terrain forcing.

\begin{figure}[htb!]
\centering
\includegraphics[width=0.91\textwidth]{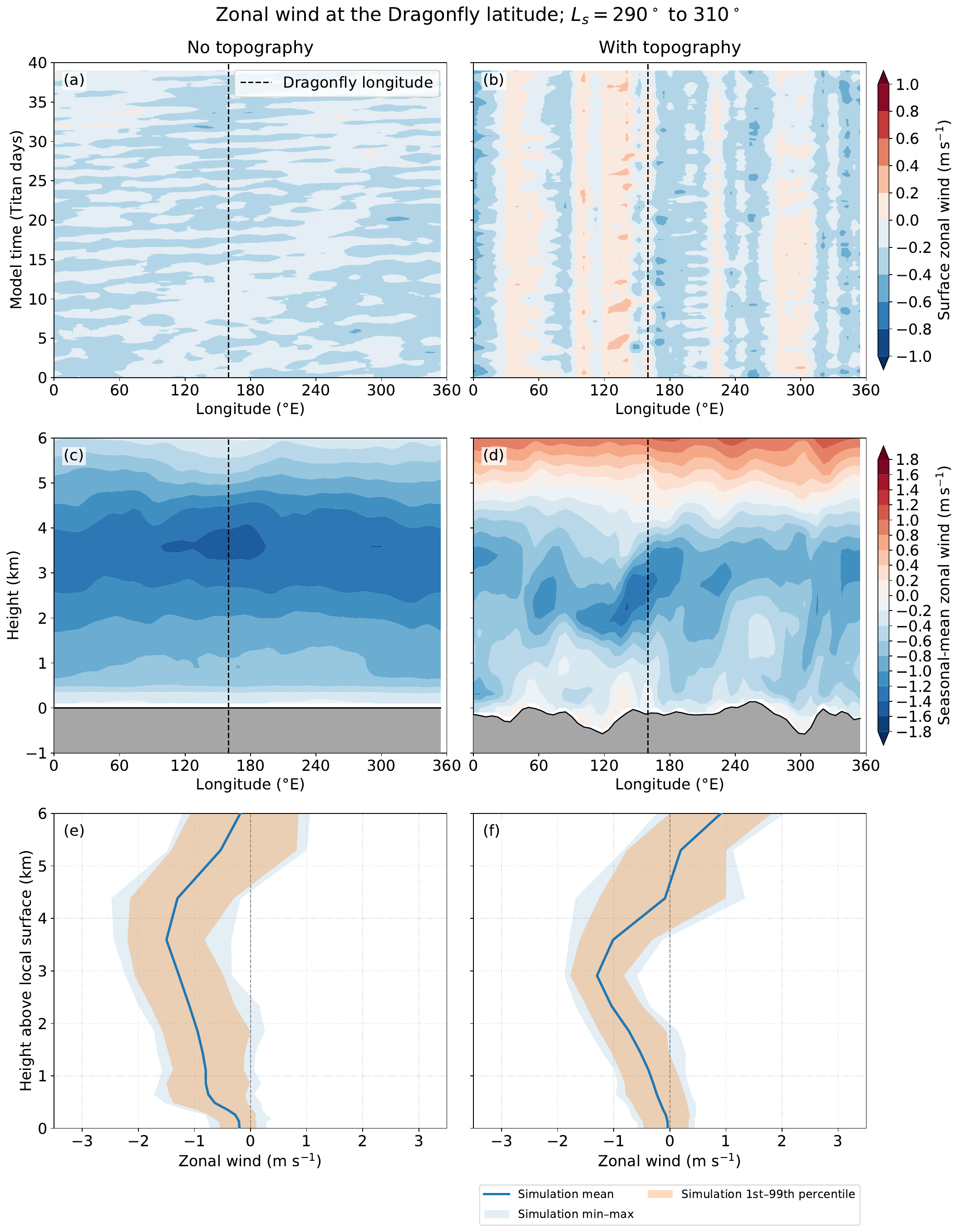}
\caption{
Similar to Figure~1, but evaluated at Dragonfly’s landing latitude of $3.7^\circ$N, represented by the model grid point centered at $2.8^\circ$N. Dragonfly’s landing longitude of $161.8^\circ$E corresponds to the same model longitude as the Huygens landing site, $163.1^\circ$E.
}
\end{figure}

\section{Conclusion}

The simulation with topography generates seasonally persistent, vertically structured stationary waves that produce localized near-surface prograde winds at the Huygens landing site, in contrast to Titan’s retrograde zonal-mean circulation. Helmholtz and spectral decompositions allow us to identify dynamically distinct contributions: a predominantly rotational $s=1$ response that provides the largest eastward anomaly within approximately the lowest $2~\mathrm{km}$, and a locally divergent $s=4$ response that partially opposes the near-surface flow. Both responses exhibit pronounced vertical variations aloft. Weaker $s=2$ and $s=3$ responses also vary with height and contribute additional eastward anomalies near the surface. Together, the $s=1$--$4$ components produce a net eastward near-surface anomaly that overcomes the retrograde background flow. The high coherence of these modes with their corresponding terrain-forcing components supports their topographic origin. For $s=4$, the coherent vertical phase tilt, gravity wave-like polarization, and positive vertical momentum flux are qualitatively consistent with a topographically forced gravity wave response. However, the propagation condition $|U|>U_{\mathrm{crit}}$ is satisfied only between approximately 2 and 4~km. The disturbance is inertially evanescent near the lower boundary, may tunnel into this limited propagation window, and becomes evanescent again above approximately 4~km before weakening near the critical-level region. The Huygens wind profile can therefore be understood as the superposition of several vertically varying, topographically forced modes on the retrograde background circulation, dominated by the rotational $s=1$ response and a gravity wave-like $s=4$ response whose vertical propagation is restricted by inertial-cutoff and critical-level effects.


The simulations also provide expectations for the atmospheric environment that may be encountered by Dragonfly. At the model grid point nearest the planned landing site, the same $s=1$ and $s=4$ stationary modes are present, although their amplitudes and vertical phases differ from those at the Huygens location. The rotational $s=1$ component again contributes eastward near-surface flow, while the more divergent $s=4$ component produces vertically alternating wind anomalies and gravity wave-like polarization. Topography consequently shifts the simulated near-surface wind toward more positive values, although its seasonal mean remains close to zero. Dragonfly may therefore encounter stationary wave-induced vertical and longitudinal wind variations similar to those inferred near Huygens, but with weaker amplitudes and a different balance between the two modes. 

Previously identified transient waves on Titan \citep{Mitchell11,BL21}  may contribute to the atmospheric variability encountered by Dragonfly and may interact with the stationary waves identified here. The two types of variability should be distinguishable through their temporal and geographic behavior: topographically forced stationary waves remain phase locked to the surface, whereas transient waves propagate relative to the terrain. The characteristic timescales, propagation properties, and underlying physical mechanisms of these transient waves will be investigated in future work.

Several limitations in our study should be noted. The T21 general circulation model has a nominal horizontal grid spacing of order $10^2~\mathrm{km}$. The simulations therefore cannot resolve individual ridges, valleys, or localized slopes smaller than a few hundred kilometers. Moreover, the Huygens and Dragonfly landing sites are separated by only approximately $680~\mathrm{km}$, corresponding to just a few model grid points. The simulated differences between the two sites should therefore be interpreted as broad regional differences in background circulation and large-scale topographic forcing rather than detailed predictions of their local environments. Consequently, short wavelength lee waves and their trapping within vertically evanescent layers are not reliably represented \citep[e.g.,][]{Durran90}. Such processes may nevertheless be important: mesoscale models have resolved localized storm outflows and lake-breeze circulations on Titan \citep{Charnay15,Rafkin20,Chatain24}, while analogous high-resolution simulations of Venus have captured trapped mountain wave responses to detailed topography \citep{Lefevre20}. Applying a comparable mesoscale framework to Titan could therefore reveal locally trapped disturbances unresolved by the present simulations \citep[e.g.,][]{Lefevre26}.

Additional uncertainty arises from Titan's incompletely measured topography. The Cassini-based global map necessarily interpolates across extensive regions lacking direct elevation constraints \citep{Corlies17}, and uncertainties in the amplitude and position of individual features could affect the phase and magnitude of the simulated stationary responses. The simulations also neglect spatial variations in surface thermal inertia, roughness, and albedo, which could alter surface energy fluxes, boundary-layer stability, and local wind amplitudes \citep{Moisan25}. Finally, we deliberately use a dry atmospheric configuration to isolate the dynamical response to topography and facilitate interpretation of the paired simulations; consequently, the model omits methane thermodynamics, cloud formation, and moist convection. These processes may modify atmospheric stability, meridional temperature gradients, and the background circulation through latent heating and gradient-wind balance \citep{Mitchell09,Lora15,Newman16,ML16}. Simulations combining topography with an interactive methane cycle will be needed to evaluate the robustness of the stationary wave response.

A further limitation concerns the dynamical interpretation of the $s=1$ response. The Helmholtz decomposition establishes that this component is predominantly rotational, but this alone does not identify a specific wave type or restoring mechanism. The local linear diagnostics do not support its interpretation as a vertically propagating inertia--gravity wave or as a standard free or vertically trapped stationary Rossby mode. We therefore describe it conservatively as a topographically forced rotational response. Further analysis of its potential-vorticity budget and meridional modal structure will be required to determine its underlying dynamics and will be pursued in future work.

Despite these limitations, our results identify large-scale topographic forcing as a plausible dynamical explanation for key features of the Huygens wind profile and provide baseline expectations for atmospheric circulation near the Dragonfly landing site. Topographically forced stationary waves should therefore be considered when interpreting past and future in-situ measurements of Titan’s atmosphere. More broadly, our results demonstrate how an individual in-situ profile can differ substantially from the zonal-mean atmospheric state, highlighting the importance of geographically localized dynamics when interpreting measurements along a single descent trajectory.

\begin{acknowledgments}

This work was funded by NASA’s Cassini Data Analysis Program under grant 80NSSC25K7446 and NASA’s Solar System Science under grant 80NSSC26K1044. B.F. was supported by the Skinner Postdoctoral Fellowship at Yale University. J.M.L. and S.R. further acknowledge support from the Dragonfly mission. ChatGPT (GPT-5.6; OpenAI) was used for grammatical refinement during preparation of the manuscript text.

The in-situ vertical profiles of zonal wind were measured by the Doppler Wind Experiment (DWE; \citealt{Bird05}), with the data available from \citet{DWE24} [\href{https://doi.org/10.17189/0qw0-e046}{doi:10.17189/0qw0-e046}]. The Helmholtz decomposition was performed using the publicly available Python package \texttt{windspharm} \citep{Dawson16}. The TAM simulation data are archived on \citet{HAN26} [\href{https://doi.org/10.5281/zenodo.23406532}{10.5281/zenodo.23406532}].

\end{acknowledgments}

\begin{contribution}


S.H., S.R., and J.M.L. conceptualized the research. S.H. performed the formal analysis and validation, and prepared the original manuscript. S.R., J.M.B., B.F., and J.M.L. reviewed and edited the manuscript. J.M.L. secured funding and provided project administration and supervision.

\end{contribution}

\restartappendixnumbering
\appendix


\section{Additional Analyses at the Huygens and Dragonfly Locations}

\begin{figure}[htbp]
  \centering
  \includegraphics[width=0.55\textwidth]{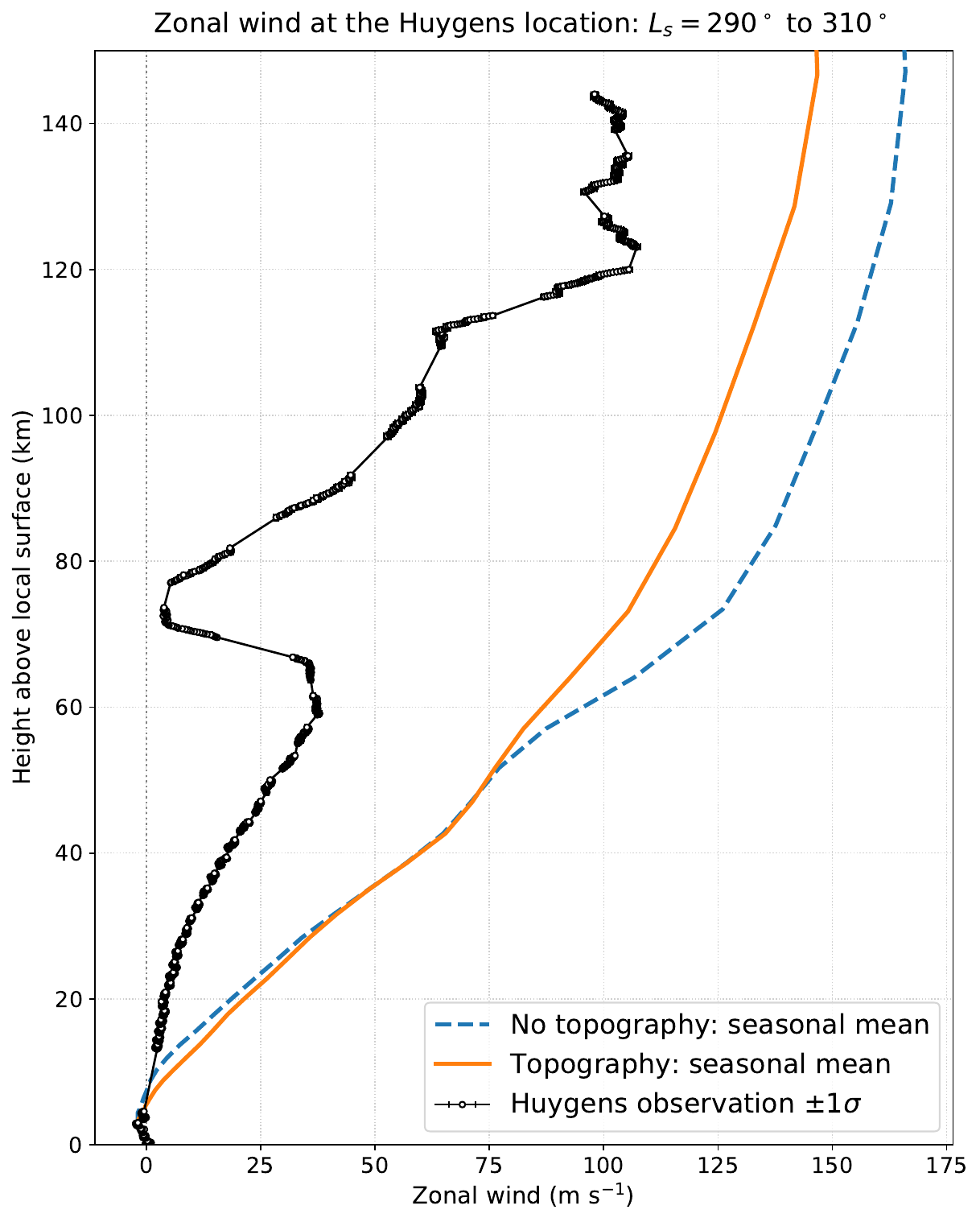}
  \caption{
  Simulated zonal wind profiles at the Huygens landing site and season from the surface to 150~km, compared with observations. 
  }
\end{figure}


\begin{figure}[htb!]
\centering
\includegraphics[width=0.95\textwidth]{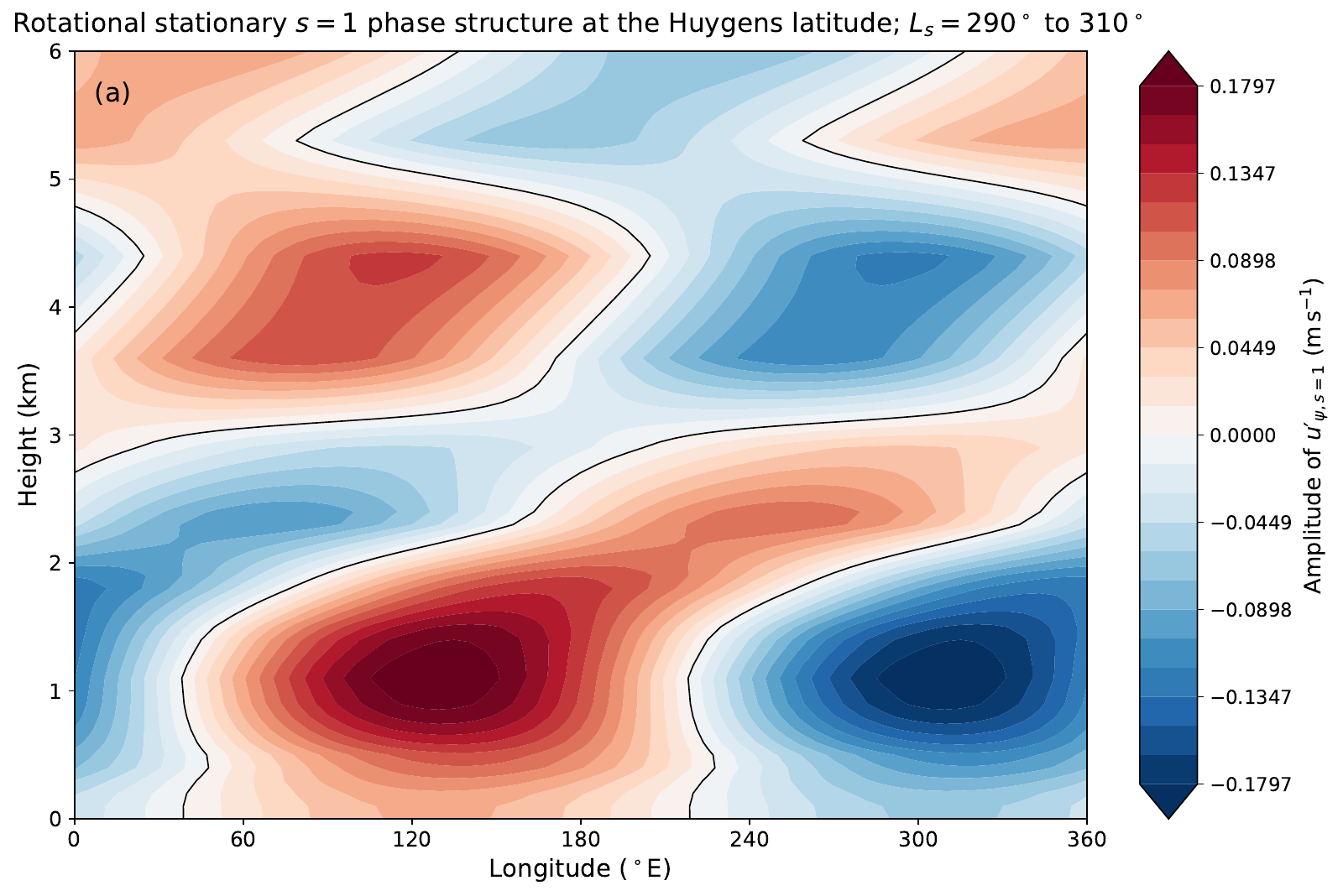}
\caption{
Longitude--height structure of the rotational component of the stationary $s=1$ zonal wind anomaly, near the Huygens latitude band during $L_s=290^\circ$--$310^\circ$. 
}
\end{figure}

\begin{figure}[htbp]
  \centering
  \includegraphics[width=0.72\textwidth]{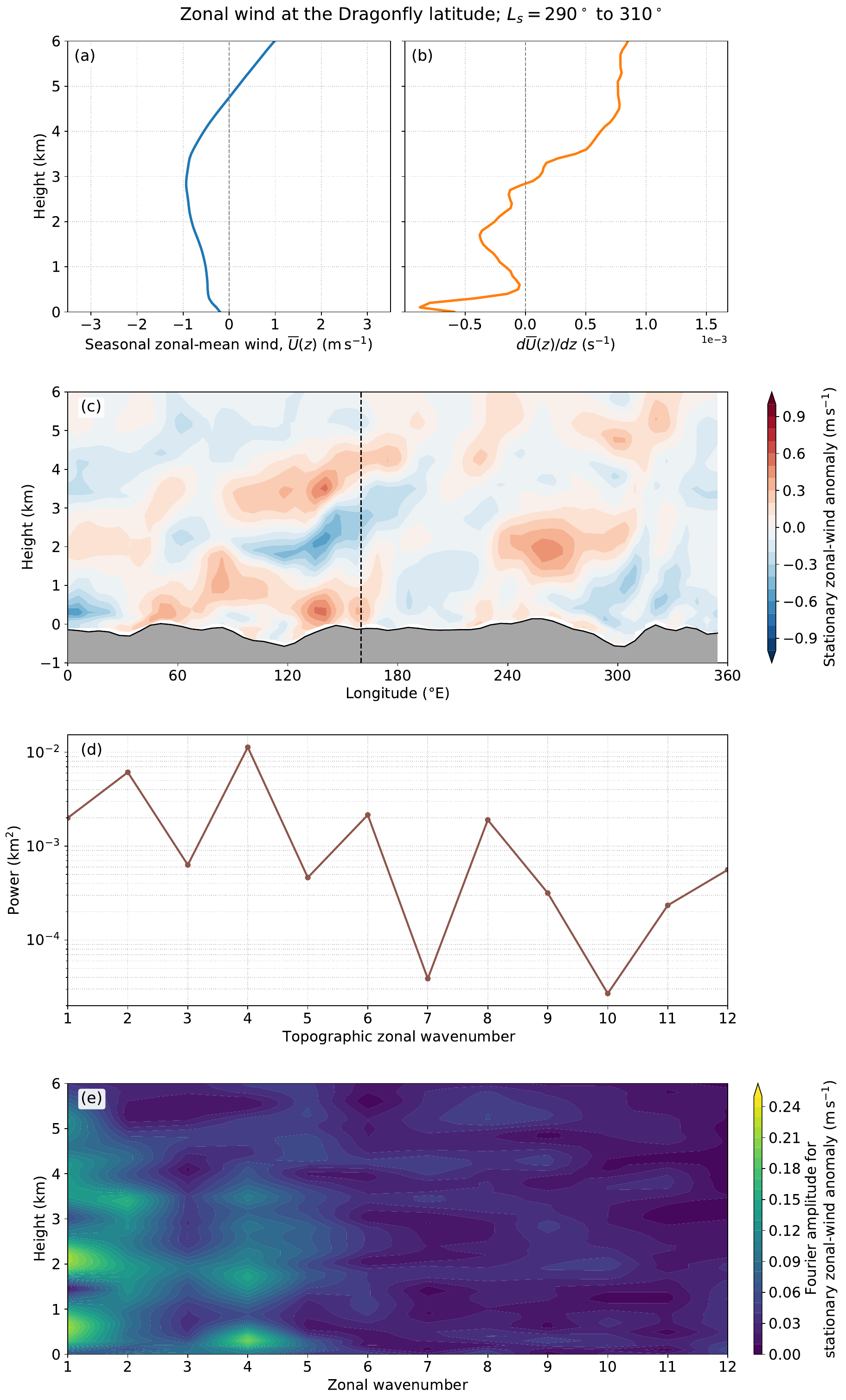}
  \caption{
  Same as Figure~2, but evaluated at the model latitude nearest to the planned Dragonfly landing site. 
  }
\end{figure}

\clearpage

\begin{figure}[htbp]
  \centering
  \includegraphics[width=1.0\textwidth]{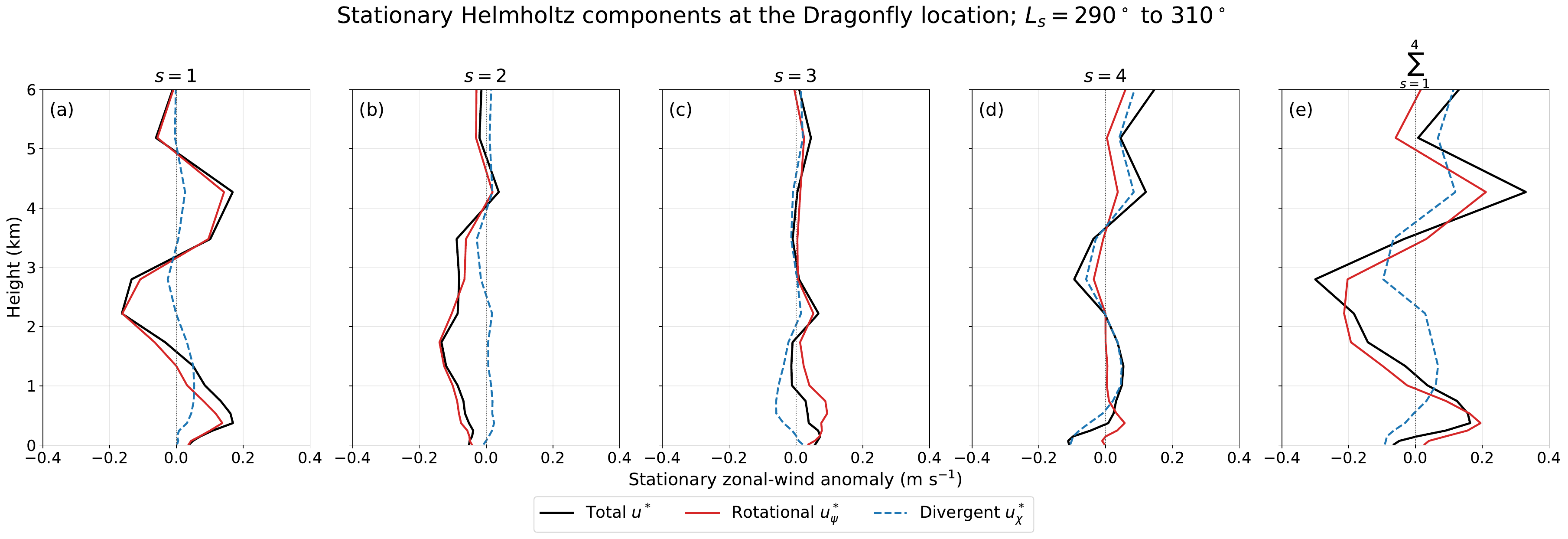}
  \caption{
  Same as Figure~3, but evaluated at the model grid nearest to the planned Dragonfly landing site. 
  }
\end{figure}

\begin{figure}[htbp]
  \centering
  \includegraphics[width=0.66\textwidth]{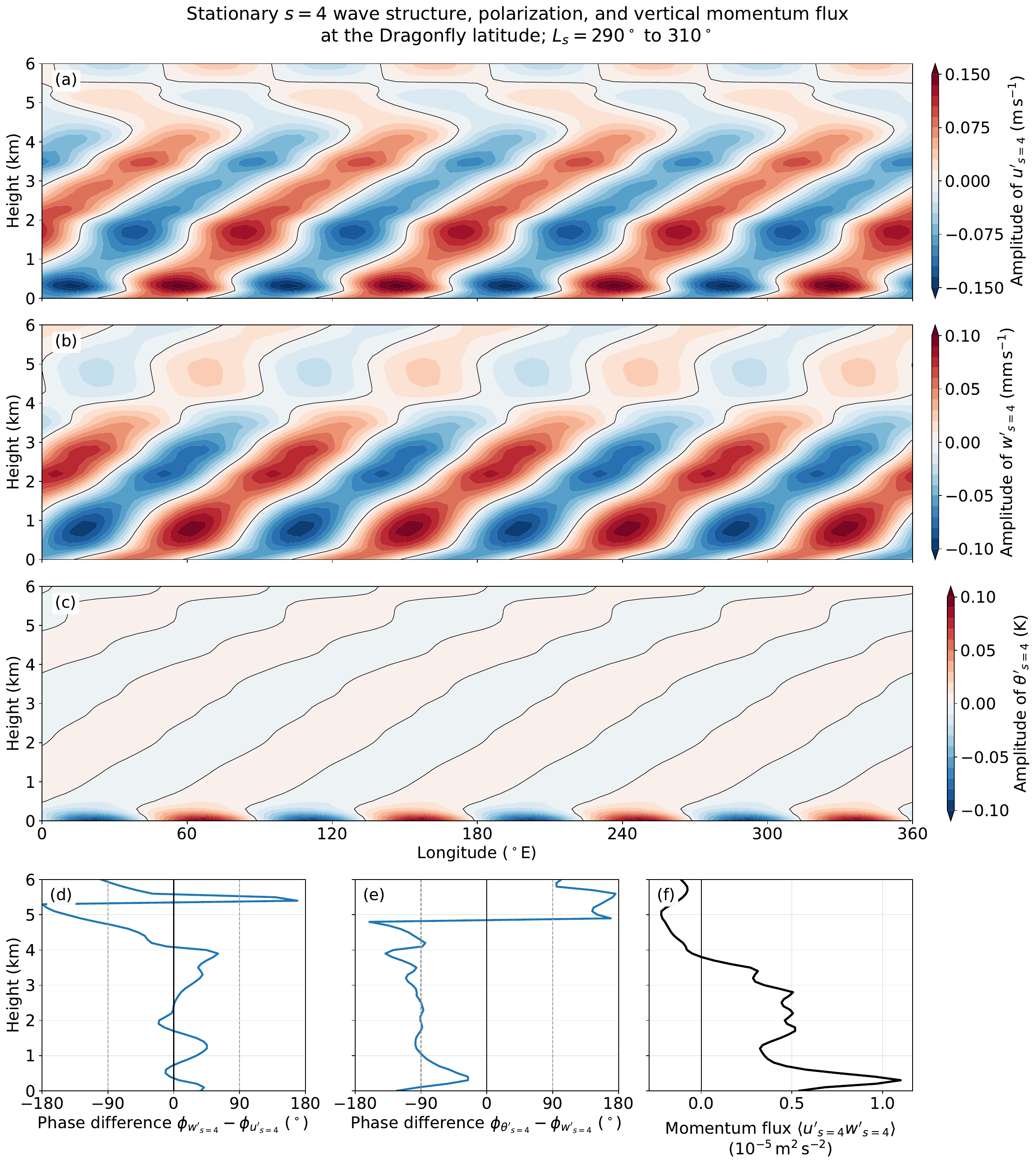}
  \caption{
  Same as Figure~4, but evaluated at the model latitude nearest to the planned Dragonfly landing site. 
  }
\end{figure}










\bibliography{Titan}{}
\bibliographystyle{aasjournalv7}

\end{document}